# A Novel Methodology for Thermal Aware Silicon Area Estimation for 2D & 3D MPSoCs

Ramya Menon C. and Vinod Pangracious

Department of Electronics & Communication Engineering, Rajagiri School of Engineering & Technology, Kochi, Kerala
ramyamenonc@gmail.com
pangracious@googlemail.com

## ABSTRACT

In a multiprocessor system on chip (MPSoC) IC the processor is one of the highest heat dissipating devices. The temperature generated in an IC may vary with floor plan of the chip. This paper proposes an integration and thermal analysis methodology to extract the peak temperature and temperature distribution of 2-dimensional and 3-dimensional multiprocessor system-on-chip. As we know the peak temperature of chip increases in 3-dimensional structures compared to 2-dimensional ones due to the reduced space in intra-layer and inter-layer components. In sub-nanometre scale technologies, it is inevitable to analysis the heat developed in individual chip to extract the temperature distribution of the entire chip. With the technology scaling in new generation ICs more and more components are integrated to a smaller area. Along with the other parameters threshold voltage is also scaled down which results in exponential increase in leakage current. This has resulted in rise in hotspot temperature value due to increase in leakage power. In this paper, we have analysed the temperature developed in an IC with four identical processors at 2.4 GHz in different floorplans. The analysis has been done for both 2D and 3D arrangements. In the 3D arrangement, a three layered structure has been considered with two Silicon layers and a thermal interface material (TIM) in between them. Based on experimental results the paper proposes a methodology to reduce the peak temperature developed in 2D and 3D integrated circuits .

## 1. INTRODUCTION

With the development of newer technologies, more devices have been integrated into a single chip. Three dimensional (3D) integration of circuit is a recent trend employed in the VLSI industry for improving the performance of the devices. The stacked devices in different layers communicate through special interconnects called Through Silicon Vias (TSV). The main drawback of 3D stacking is the considerable rise in the temperature which can eventually lead to the damage of the device. A thermal aware architecture is very essential to solve this problem. The first step in the design of such architecture is to find out the heat dissipating devices and analyse the temperature developed in different floorplans. Here we have done experiments to find out the extent to which the temperature can shoot up in 2D and 3D IC with four processors with the help of Hotspot simulator. This simulator takes the files with floorplan and power of devices as inputs. All the processors used in our experiments were identical. Each of them consumed 50.9 Watt power in the active mode.

## 2. PROBLEM FORMULATION

Now-a-days, high importance is given to three dimensional integration of the circuits. Experiments have been done by Annmol Cherian et al to analyze the static power dissipation in 3D stacked memories for modelling inter-layer temperatures [4]. The authors have studied various second order effects of the transistors and static power dissipation of memory is analyzed. Gabriel H. Loh conducted special studies on improving speed of 3D DRAM by changing the organization [8]. Loh proposed the idea of combining Vector Bloom Filter with dynamic MSHR capacity tuning to address L2 miss handling architecture (MHA). Optimized architectures for 3D stacked memories to overcome the memory bandwidth problem have been proposed by Dong Hyuk Woo et al [15]. A new technique called SMART-3D has been proposed to increase the speed of the memory. This technique reduces the energy consumption in L2 caches and 3D DRAM. In short, lots of research works are going on in the field of 3D integration of memories. The heat generated in a device mainly depends on the power that it dissipates. Compared to memories the power consumed by the processors is very high. Hence the processors are the highest heat dissipating devices and more importance should be given to the processor temperature analysis. It is in this context, our experiments with 2D & 3D stacked processors gain relevance. Moreover, 3D integration has got many challenges associated with it. 3D circuits are more susceptible to noise and stress failures due to rise in temperature. As this affects the reliability of the chip very much a thermal aware floorplan is of prime importance.

## 3. 2-DIMENSIONAL DESIGN AND ANALYSIS

Four simulations were carried out with 2D floorplan. The set up for all the four simulations were same. A square shaped Silicon substrate with width 0.016m was taken. The total Silicon area was .000256 $m^2$ and the area occupied by a single processor was .000016 $m^2$. In the first simulation the heat dissipated by a single processor has been studied [3]. This work has been done previously by Ramya Menon C. et. al. to study the affect of area occupied by the processor on Silicon. In each new generation, this area is scaled down. The temperature and area maintains an inverse relationship, as one (area) decreases, the other (temperature) increases. The next three simulations were carried out with four processors in different floorplans. In the second simulation all the processors were placed adjacent to each other. In the third simulation all the processors were placed diagonally and in the fourth one, the processors were placed at the four corners of the substrate.

### 3.1. Design with Single Processor to Identify Hotspot Location and Temperature

A single processor was placed in the bottom right corner of the substrate [3]. The highest temperature location (hotspot) was in the processor area. The peak temperature obtained was 354.96K and the lowest was 323.19K. The thermal profile obtained is as shown in fig.1. The hotspot is in deep red color and the coolest region is in dark blue color. As the temperature increases the color gradually changes in the order dark blue, light blue, green, yellow, orange and finally red.

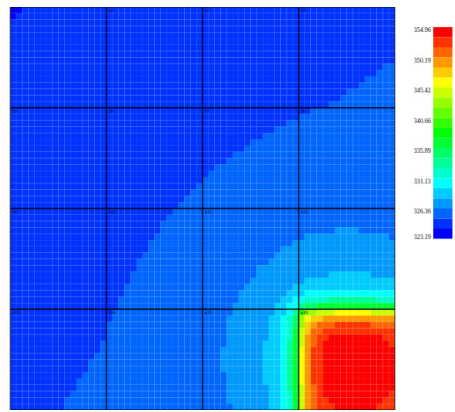

Figure 1. Thermal profile of design with single processor

## 3.2. Design with Adjacent Processors

Here, the processors were placed adjacent to each other in a single line. The thermal profile obtained is shown in fig.2. The highest temperature obtained was 380.65K and the lowest temperature obtained was 339.97K. The hotspot was mainly located in the two processors located in the middle.

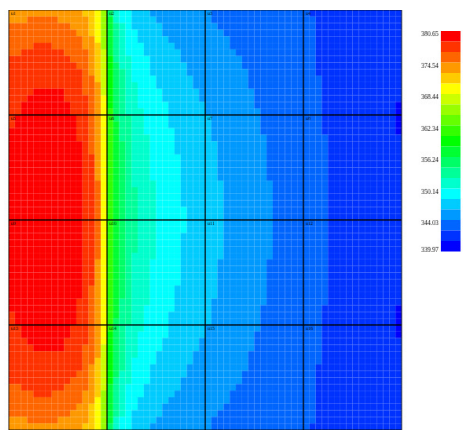

Figure 2. Thermal profile with adjacent processors

## 3.3. Design with Diagonally Placed Processors

The four processors were placed diagonally in the substrate. The thermal profile obtained is shown in fig.3. Here also the hotspot developed in the two processors placed in the middle. The highest temperature developed was 376.21K and the lowest was 343.23K.

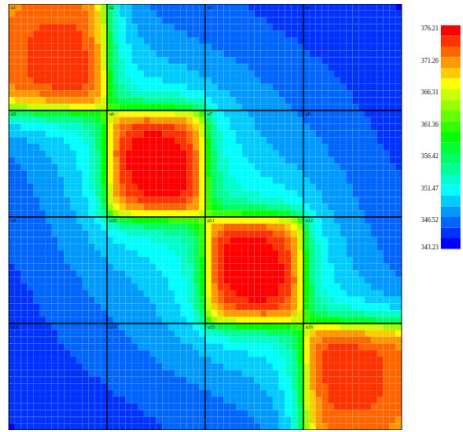

Figure 3. Thermal profile of diagonal design

### 3.4. Design with Best Balanced Temperature

In this design, the four processors were placed at the corners of the substrate. This was an arrangement in which the processors were placed at maximum distance from each other. The thermal profile is shown in fig.4. The hotspot developed in each of the processor units. The hotspot temperature was noted as 372.76K and the least temperature was 344.52K.

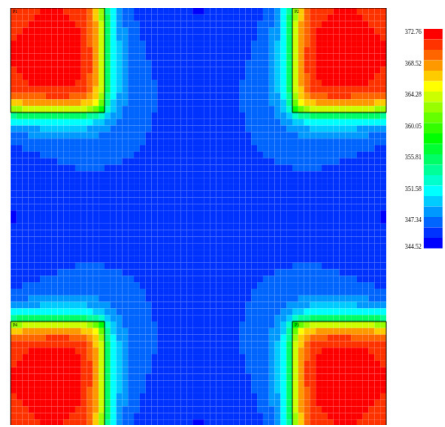

Figure 4. Thermal profile of the best balanced design

### 3.5. Result Analysis of 2D Design

The highest and lowest temperature obtained in each of the 2D design is shown in table.1. There were three 2D simulations with 4 processors- simulation 2 (EXP 2), simulation 3 (EXP 3) & simulation 4 (EXP 4). Fig.5 shows the plot of highest and lowest temperatures in each of these designs. The x-axis shows simulation number and the y-axis shows the temperature (T) in Kelvin (K).

Table.1 Highest and lowest temperatures of 2D experiments

| **Experiment No:** | **Highest Temperature (K)** | **Lowest Temperature (K)** |
|---|---|---|
| 1 | 354.96 | 323.19 |
| 2 | 380.65 | 339.97 |
| 3 | 376.21 | 343.23 |
| 4 | 372.76 | 344.52 |

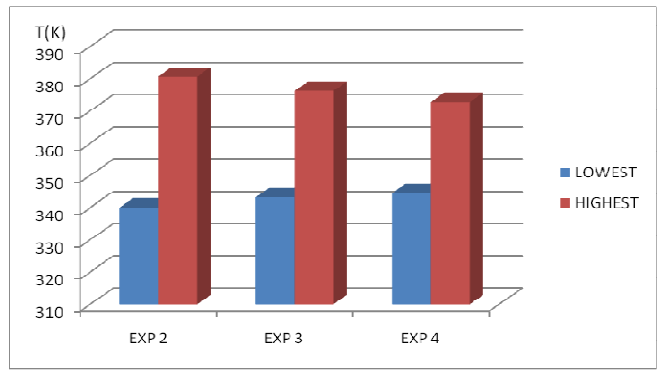

Figure 5. Plot of highest and lowest temperatures

## 4. 3-DIMENSIONAL INTEGRATION AND THERMAL ANALYSIS

In the second phase simulations, the four processors were three dimensionally stacked in three layers. The bottom layer (layer 0) was a Silicon layer with two processors. The intermediate layer (layer 1) was the thermal interface material (TIM). The top layer (layer 2) was another Silicon layer with two processors. In the different simulations the floorplans of the Silicon layers was varied. TIM used here was epoxy based resin.

### 4.1. Design with Direct Vertical Stacking

The floorplans of different layers in this design were as shown in the figure 6. The shaded region represents the processors in layers 0 and 2. The two processors in the layer 2 were vertically above those in layer 0. The aim of this was to find maximum temperature generated in the processor units.

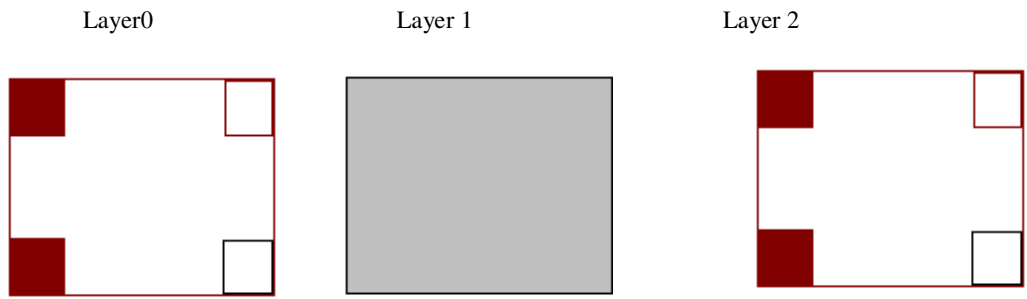

Figure 6. Floorplan with direct vertical stacking

The thermal profile obtained by taking a single 2D layer (layer 0) of this design without layer 1 and layer 2 above it is as shown in figure 7.The hotspot was developed in processor units. The hotspot temperature was 361.29K and the least temperature noted was 329.01K.

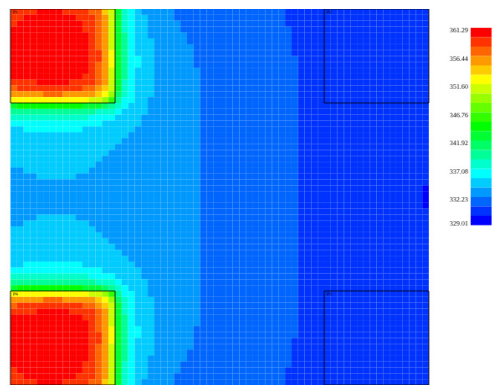

Figure 7. Thermal profile of layer 0

After stacking all the three layers one above the other, the temperature developed in each of the units increased. The result is shown in table 2. The processors in layer 0 had highest temperatures, 392.72K. TIM in between the processors had a temperature of 388.07K.

Table 2. Result of direct vertical stacking of processors

| Peak temperature observed in layer 0 processors(K) | Peak temperature observed in layer 2 processors(K) | Peak temperature observed in layer 1 TIM (K) | Lowest temperature developed (K) |
|---|---|---|---|
| 392.72 | 372.56 | 388.07 | 341.25 (in all the 3 layers) |

### 4.2. Design with Indirect Vertical Stacking

In this design, the floorplans of different layers were as shown in the figure 8. The shaded region represents the processors in layers 0 and 2. Layer0 in this experiment had the same floorplan as the layer 0 of fig.6. In layer 2 the processors were placed in the other two corners so that there was no processor exactly above the other.

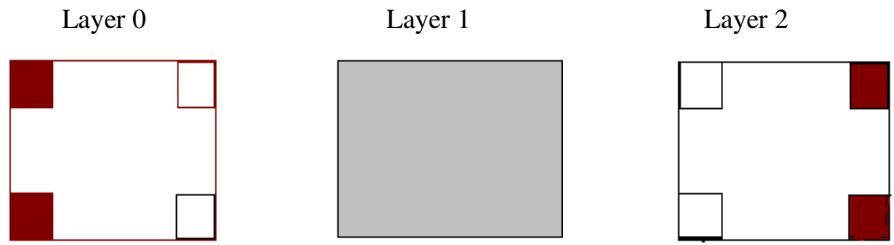

Figure 8. Floorplan with indirect vertical stacking

After stacking all the three layers one above the other, the temperature developed in each of the units were analysed. The result is shown in table 3. The processors in layer 0 had highest temperatures, 377.22K. The temperature developed in the processors in layer 2 was 356.98K.

Table 3. Result of indirect vertical stacking of processors

| Peak temperature observed in layer 0 processors(K) | Peak temperature observed in layer 2 processors(K) | Peak temperature observed in layer 1 TIM (K) | Lowest temperature developed (K) |
|---|---|---|---|
| 377.22 | 356.98 | 372.51 | 356.76 (in layer 0) |

### 4.3. Design with Direct Diagonal Vertical Stacking

The floorplan of this design is as shown in figure 9. The processors were placed in the diagonally opposite corners in each Silicon layer. The processors in layer 2 were exactly above the processors in layer 0.

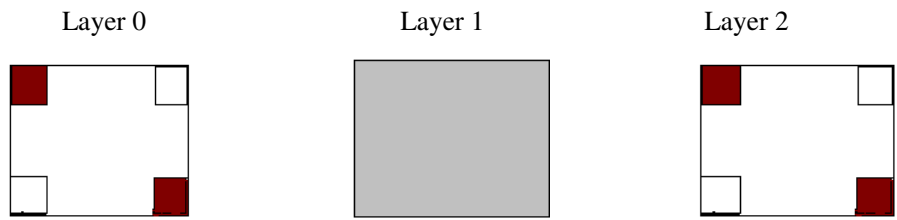

Figure 9.Floorplan with direct diagonal vertical stacking

The thermal profile obtained by taking a single 2D layer (layer 0) of this design without layer 1 and layer 2 above it is as shown in fig.10. Hotspots were developed in the centre of the processor units. The hotspot temperature was 360.16K and the least temperature developed was 330.54K.

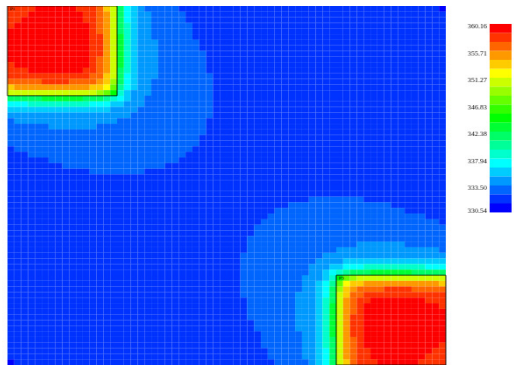

Figure 10. Layer 0 thermal profile with diagonal placement

The temperature values were analysed after stacking all the three layers one above the other. The result is shown in table 4. The processors in layer 0 developed a temperature of 390.57K and that in the layer 2 developed a temperature of 370.41K. TIM in between processors had a temperature of 385.92K.

Table 4. Result of direct diagonal vertical stacking of processors

| Peak temperature observed in layer 0 processors(K) | Peak temperature observed in layer 2 processors(K) | Peak temperature observed in layer 1 TIM (K) | Lowest temperature developed (K) |
| --- | --- | --- | --- |
| 390.57 | 385.92 | 372.51 | 343.40 (in all 3 layers) |

## 4.4. Design with Indirect Diagonal Vertical Stacking

The floorplan is as shown in fig.11. Here also the processors were placed in diagonally opposite corners of the silicon layers. But none of the processors were exactly above the other in different layers.

Layer 0                     Layer 1                     Layer 2

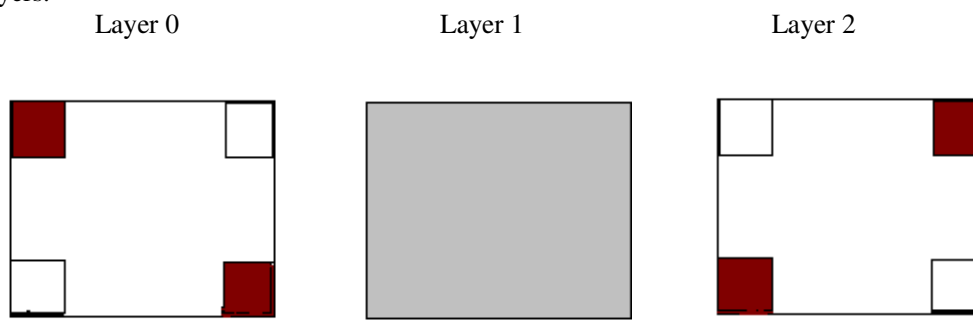

Figure 11. Floorplan with indirect diagonal vertical stacking

The result obtained after stacking all the three layers one above the other is shown in table 5. The processors in the layer 0 attained a temperature of 377.22K and those in layer 2 attained a temperature of 356.98K.

Table 5. Result of indirect diagonal vertical stacking of processors

| Peak temperature observed in layer 0 processors(K) | Peak temperature observed in layer 2 processors(K) | Peak temperature observed in layer 1 TIM (K) | Lowest temperature developed (K) |
| --- | --- | --- | --- |
| 377.22 | 356.98 | 356.76 | 343.40 (in layer 0) |

### 4.5. Result Analysis of 3D Experiments

The temperature developed in the processors in different layers of 3D experiments is shown in table 6. Figure 12 shows the plot of temperatures given in table 6.

Table 6. Temperature Values of 3D experiments

| Experiment No: | Layer 0 Processors | Layer 2 Processors |
| --- | --- | --- |
| 1 | 392.72 | 372.56 |
| 2 | 377.22 | 356.98 |
| 3 | 390.57 | 370.41 |
| 4 | 377.22 | 356.98 |

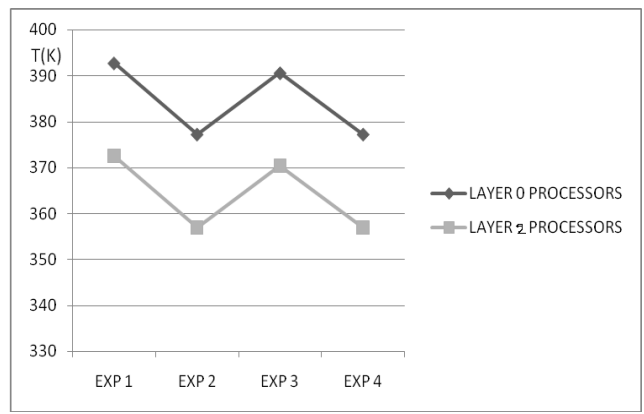

Figure 12. Temperature Plot of 3D simulations

## 5. CONCLUSIONS & FUTURE WORK

As the number of processors increases the hotspot temperature increases. From the 2D experiments arranging the neighbouring processors diagonally is a thermal aware design compared to adjacent placement. The processors should be far away from each other as much as possible. From the 3D experiments, a processor should not be stacked vertically above the other. In all the 3D experiments the bottom layer (layer 0) processors developed higher temperatures when compared to layer 2 processors. The interesting fact noted was that the layer 0 and layer 2 temperatures differed approximately by 20K in all the experiments. In the experiments with one processor vertically stacked above the other, the TIM in between them had a temperature higher than that of layer 2 processor but lower than that of layer 0 processor. The 3D experiments in which the processors were not vertically stacked above the other (EXP 2 & EXP 4) gave same processor temperature values. Now that the peak temperature values and the floorplans that minimizes these have been found out, our next step is to work on the thermal management solutions for 3D integrated circuits which can further reduce the hotspot temperature. We will also work on TSV modelling. As TSV penetrates through the various Silicon layers the temperature analysis will become more complicated. Thermal Through Silicon Via (TTSV) will also be introduced and analysed.

**Authors**

Ramya Menon. C completed B. Tech degree in Electronics and Communication Engineering, from S.C.T College of Engineering, affiliated to Kerala University, Kerala. She is presently pursuing M.Tech in VLSI and Embedded Systems, at Rajagiri School of Engineering and Technology, India. She is the author of the paper titled "A Comparative Study of Placement of Processor on Silicon and Thermal Analysis", which has been accepted in CCSIT-2012 to be held in Bangalore. The project currently working on is thermal modeling & 3D stacking of integrated circuits.

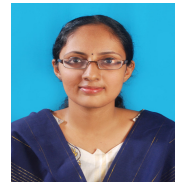

Vinod Pangracious received the B.Tech degree in Electronics Engineering from Cochin University of Kerala in 1995 and M.Tech degree from IIT Bombay India in 2000. Currently pursuing PhD at University of Pierre and Marie Curie Paris France. He is currently an Associate Professor at Rajagiri School of Engineering & Technology India. He has authored and co-authored 10 publications in these areas. He is an internationally experienced electronics engineering professional with extensive expertise in memory design, high speed digital circuit design, logic library development, verification, test and characterization. His research interest focus on design methodologies for integrated systems, including thermal management technique for multiprocessor system on chip, novel nanoscale architectures for logic and memories, dynamic memory, 3D integration and manufacturing technologies.

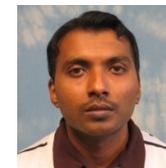